\documentclass[amssymb,pra,twocolumn,notitlepage,superscriptaddress,longbibliography]{revtex4-2}

\usepackage{graphicx}
\usepackage{bm}
\usepackage{amsmath}
\usepackage{color}
\newcommand\revise[1]{#1}

\usepackage{braket}
\usepackage[unicode=true,pdfusetitle,
 bookmarks=true,bookmarksnumbered=false,bookmarksopen=false,
 breaklinks=false,pdfborder={0 0 0},backref=false,colorlinks=true]
 {hyperref}

\begin{document}

\title{Unifying Charge-Learnability Transitions in Monitored \(U(1)\)-Symmetric Quantum Circuits through Informational Power of Local Measurement}

\author{Yi-Fan Gong}
 \affiliation{Key Laboratory of Atomic and Subatomic Structure and Quantum Control (Ministry of Education), Guangdong Basic Research Center of Excellence for Structure and Fundamental Interactions of Matter, and School of Physics, South China Normal University, Guangzhou 510006, China}

\author{Dan-Bo Zhang}
 \altaffiliation[Contact author: ]{dbzhang@m.scnu.edu.cn}
 \affiliation{Key Laboratory of Atomic and Subatomic Structure and Quantum Control (Ministry of Education), Guangdong Basic Research Center of Excellence for Structure and Fundamental Interactions of Matter, and School of Physics, South China Normal University, Guangzhou 510006, China}
 \affiliation{Guangdong Provincial Key Laboratory of Quantum Engineering and Quantum Materials, Guangdong-Hong Kong Joint Laboratory of Quantum Matter, and Frontier Research Institute for Physics, South China Normal University, Guangzhou 510006, China}
\date{\today}

\begin{abstract}
Charge-learnability transitions in monitored symmetric quantum circuits reveal how local measurement records acquire sufficient information to infer a conserved charge. Here we extend charge learnability to probabilistic weak measurements, for which the measurement probability and measurement strength are independently tunable. We find that the learnability phase boundary is organized by the informational power of local measurement. We further introduce cross entropy as a label-sensitive diagnostic that distinguishes unbiased, biased, and antibiased decoder variants. Finally, the exact record--label mutual information provides a decoder-independent benchmark for the information fundamentally available for charge inference. Our results establish informational power of local measurement as a unifying principle for charge learnability under general monitoring protocols.
\end{abstract}

\maketitle

\section{Introduction}
Monitored quantum circuits provide an experimentally accessible setting in which measurement acts as a genuine dynamical ingredient rather than as a passive probe. The competition between unitary scrambling and local measurements can drive measurement-induced phase transitions, most notably between volume-law and area-law entanglement scaling \cite{Li2018,Li2019,Chan2019,Skinner2019,Bao2020,Jian2020,Choi2020,GullansHuse2020Scalable,GullansHuse2020,Nahum2021,Zabalo2022,Li2021CFT,Ippoliti2021,Fan2021,Sang2021,Turkeshi2020}. Related developments have explored purification, free-fermion, long-range, and other extensions of monitored dynamics \cite{LuntPal2020,Alberton2021,Block2022,Minato2022,VanRegemortel2021}. Conservation laws enrich this phenomenology further. In circuits with a conserved $U(1)$ charge, measurements not only affect entanglement and purification, but also acquire information about globally conserved quantities, leading to charge-sharpening transitions, charge-fluctuation criticality, and related symmetry-resolved phenomena \cite{Khemani2018,Rakovszky2018,Agrawal2022,BarrattFieldTheory2022,OshimaFuji2023,Bao2021Symmetry,Lavasani2021,HanChen2022}.

Measurement records provide classical data for learning quantum systems. Remarkably, in monitored dynamics the measurements also reshape the trajectory through backaction and collapse and thus provide a new perspective on learning quantum systems. Recent studies identify learnability transitions in settings including inference of conserved charges from local measurement outcomes~\cite{BarrattLearnability2022}, eavesdropper reconstruction using classical shadows~\cite{IppolitiKhemani2024Shadows}, the observation of measurement collapse through learnability~\cite{Utkarsh_PRX_2024}, and mixed-state charge learnability in noisy monitored dynamics~\cite{Hansveer_PRB_2026}.
Collectively, they show that correlations encoded in a spacetime measurement record can undergo a sharp transition from being insufficient at low-density measurements to being asymptotically sufficient at high-density measurements for reconstructing the charge. Most existing formulations, however, focus on probabilistic projective measurements. In this case, each realized measurement is maximally informative in the chosen basis and thus is quite sharp, so the rate at which information is acquired is controlled primarily by the measurement probability.  Meanwhile, more general measurement protocols have been extensively studied in the context of measurement-induced entanglement transitions, including weak and generalized  measurements\cite{Noel2022,Koh2023,Hoke2023,Szyniszewski2019,Szyniszewski2020,FujiAshida2020,VanRegemortel2021,Aziz2024,MochizukiHamazaki2025}. For such measurements, however, the measurement probability and measurement strength are distinct control parameters. Whether a learnability transition persists when these two quantities are separated, and what variable governs the transition, remain open questions. This motivates us to study charge learnability in monitored random circuits with probabilistic weak measurements, where the measurement probability and measurement strength can be tuned independently.

In this work, we extend charge learnability to probabilistic weak measurements. We show that the learnability phase boundary is governed by the local informational power of the measurement, thereby unifying probabilistic projective measurements, deterministic weak measurements, and general probabilistic weak measurements within a single phase diagram. We further introduce cross entropy as a learnability diagnostic that clearly distinguishes unbiased, biased, and antibiased decoders. We show that the intrinsic limit of learnability is characterized by the mutual information between the charge and the measurement record, providing an intrinsic benchmark for charge inference. Taken together, our results establish measurement informational power as the organizing principle of the learnability phase diagram and provide a unified framework for comparing both monitoring protocols and decoder performance.

The paper is organized as follows. In Sec.~\ref{SecII}, we introduce charge learning in monitored $U(1)$-symmetric quantum circuits under probabilistic weak measurements and define measurement informational power. In Sec.~\ref{SecIII}, we determine the learnability phase boundary numerically and compare the three decoder variants. In Sec.~\ref{sec:mi}, we analyze the decoder-independent limit set by the record--label mutual information. We conclude in Sec.~\ref{SecIV}. 

\section{Charge learnability under probabilistic measurements}
~\label{SecII}
\begin{figure}[t]
\centering
\includegraphics[width=\columnwidth]{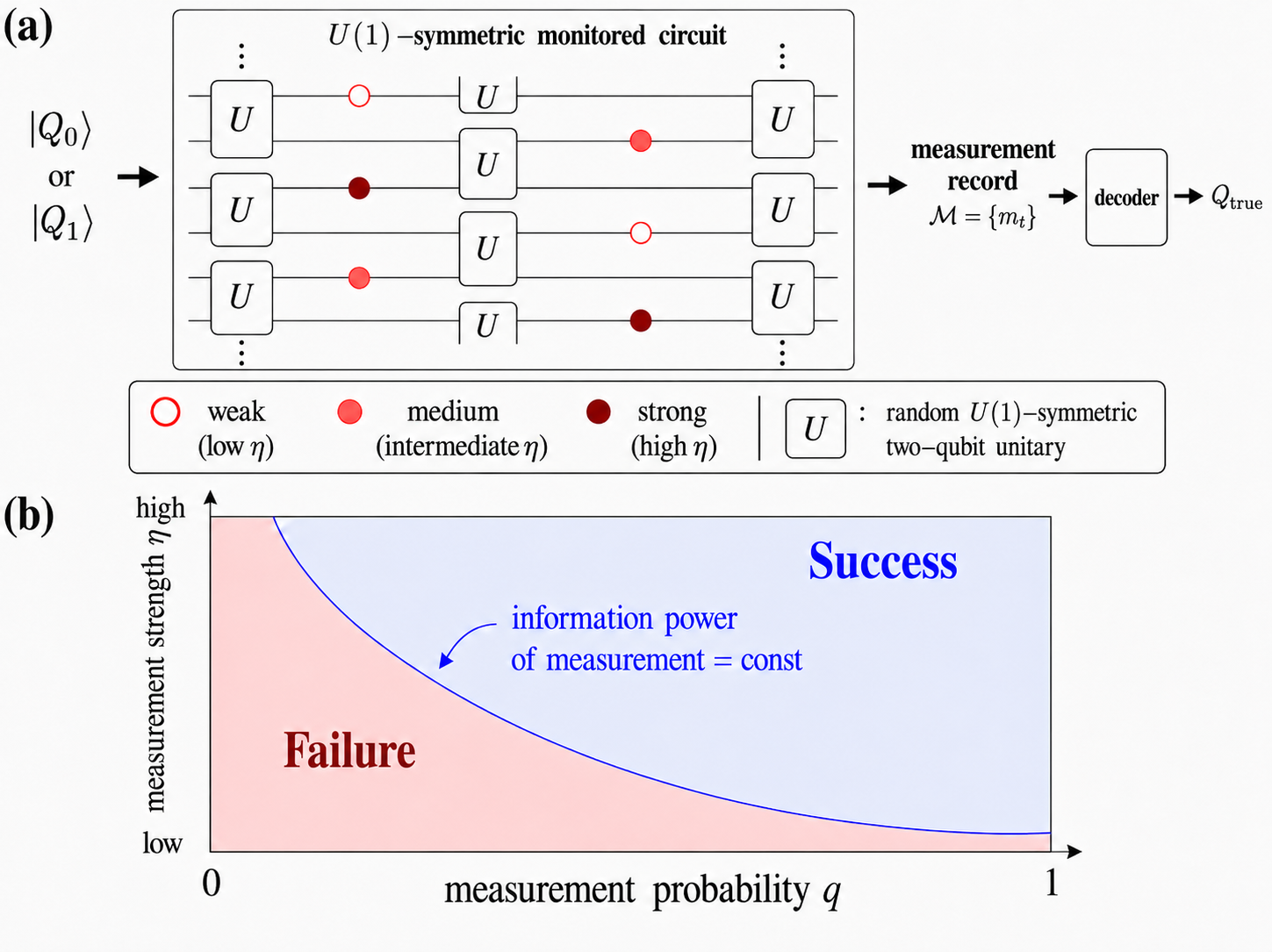}
\caption{
Schematic of charge learnability in a monitored \(U(1)\)-symmetric quantum circuit and its phase diagram in terms of the measurement probability \(q\) and measurement strength \(\eta\). (a) The circuit is initialized in one of two neighboring charge sectors. Random charge-conserving two-qubit gates are arranged in alternating even- and odd-bond layers, each followed by a probabilistic local weak-measurement layer. The resulting spacetime record is processed by a decoder to infer the hidden charge label. (b) A transition boundary separates the failure and success regimes and is organized by the measurement informational power.
}
\label{fig:schematic}
\end{figure}

We formulate charge learnability in a $U(1)$-symmetric circuit with probabilistic weak measurements. We then introduce the symmetric-exclusion-process~(SEP) decoder, its variants and finite-size diagnostics, and the informational power used to compare monitoring protocols. Figure~\ref{fig:schematic} summarizes the circuit, decoding task, and charge-learnability phase diagram. 

\subsection{Monitored quantum circuit and learning protocol}
Following Ref.~\cite{BarrattLearnability2022}, we consider charge inference in a one-dimensional $U(1)$-symmetric circuit with a generalized local readout channel. A chain of \(L\) qubits evolves under a brickwork circuit of random charge-conserving two-qubit gates~\cite{Agrawal2022}. The conserved total charge is
\begin{equation}
Q=\sum_i n_i,
\qquad
n_i=\frac{z_i+1}{2},
\label{eq:charge}
\end{equation}
where $z_i=\pm1$ are the eigenvalues of the Pauli operator $Z$. 
The hidden binary label specifies one of the neighboring charge sectors
\(Q_0=L/2\) and \(Q_1=L/2-1\), chosen with equal prior probabilities. The observer receives the complete spacetime measurement record \(\mathcal M\) generated by a single monitored trajectory and infers the hidden charge label. 

At each monitored spacetime point, a local measurement is independently applied with probability \(q\), which controls the spacetime density of recorded outcomes. Conditioned on its application, the local charge is weakly measured using the Kraus operators
\begin{equation}
M_{\pm}(\eta)=\frac{I\pm \eta Z}{\sqrt{2(1+\eta^2)}},
\label{eq:kraus_compact}
\end{equation}
where $I$ is the identity and \(\eta\) is the measurement strength. The projective limit is recovered at \(\eta=1\), whereas \(\eta\to0\) gives a noninformative readout. Thus \((q,\eta)\) independently controls the measurement probability and the information carried by each realized outcome. The complete spacetime record is \(\mathcal M=\{m_{i,t}\}\), where \(m_{i,t}\in\{+,-,\varnothing\}\) denotes the local record at site $i$ and layer $t$, and \(\varnothing\) denotes that no measurement was performed.

The inference task is implemented by a Bayesian decoder that identifies the candidate charge sector most consistent with the spacetime record. We use a classical learner that represents the coarse-grained charge dynamics by a one-dimensional symmetric exclusion process~\cite{BarrattLearnability2022}. 

For a candidate sector \(Q\), we initialize a classical probability vector \(|Q)\) uniformly over computational-basis configurations with total charge \(Q\). Within the decoder, the charge dynamics are represented by an effective SEP transfer matrix. In the Haar-averaged unbiased decoder, the two-site update acts on the local charge basis \(\{|00\rangle,|01\rangle,|10\rangle,|11\rangle\}\) as
\begin{equation}
T_{\mathrm{SEP}}
=
\begin{pmatrix}
1 & 0 & 0 & 0\\
0 & 1/2 & 1/2 & 0\\
0 & 1/2 & 1/2 & 0\\
0 & 0 & 0 & 1
\end{pmatrix}.
\label{eq:sep_transfer_unbiased}
\end{equation}
The states \(|00\rangle\) and \(|11\rangle\) are fixed by charge conservation, while the central block describes the stochastic exchange of \(|01\rangle\) and \(|10\rangle\).

The likelihood of an observed record $\mathcal M$ depends on the measurement model. For a projective measurement, the outcomes $+$ and $-$ uniquely determine the local charge. For a weak measurement, both charge values remain compatible with either outcome but acquire different likelihood weights. A recorded outcome \(m=\pm1\) updates the probability vector through the diagonal matrix
\begin{equation}
R(\eta;m)
=
\begin{pmatrix}
\frac{(1+m\eta)^2}{2(1+\eta^2)} & 0 \\
0 & \frac{(1-m\eta)^2}{2(1+\eta^2)}
\end{pmatrix}.
\label{eq:weak_measurement_weight}
\end{equation}
No weight is applied when \(m_{i,t}=\varnothing\). Weak measurements therefore replace the hard consistency constraints of projective measurements by positive outcome-dependent weights. In the projective limit \(\eta=1\), configurations incompatible with a recorded outcome receive zero likelihood. 

For decoder variant \(d\), let \(\mathcal T_d(\mathcal M;\eta)\) denote the classical transfer operator obtained by composing the SEP updates and weak-measurement weights over all circuit layers. The likelihood of record $\mathcal M$ conditioned on candidate sector $Q$ is
\begin{equation}
\mathcal L_d(\mathcal M|Q)
=
(1|\,\mathcal T_d(\mathcal M;\eta)\,|Q),
\label{eq:sep_likelihood}
\end{equation}
where \((1|\) denotes the unnormalized sum over all computational-basis configurations. For the balanced binary task, the candidate sectors $Q_0$ and $Q_1$ have equal prior probabilities, and the posterior is
\begin{equation}
\hat P_d(Q|\mathcal M)
=
\frac{\mathcal L_d(\mathcal M|Q)}
{\mathcal L_d(\mathcal M|Q_0)+\mathcal L_d(\mathcal M|Q_1)}.
\label{eq:decoder_posterior}
\end{equation}
The predicted label is the maximum-posterior sector, while the full posterior distribution is retained for the finite-size diagnostics defined below.

Equation~\eqref{eq:sep_likelihood} is a tensor-network contraction of a one-dimensional classical transfer problem. Direct evolution of the full probability vector scales exponentially with system size; instead, the sequence of two-site SEP matrices and diagonal measurement weights is contracted using matrix-product-state and TEBD methods, following Ref.~\cite{BarrattLearnability2022}. The physical record \(\mathcal M\) is generated by a monitored quantum trajectory, whereas the decoder evaluates a classical partition sum over charge histories compatible with that record.

\revise{For the present sizes \(L=6,8,10\), exact state-vector propagation would also be feasible for generating the physical measurement records. The tensor-network treatment is retained for the SEP decoder because its likelihood evaluation has the structure of a one-dimensional transfer-matrix contraction, following Ref.~\cite{BarrattLearnability2022}.}

To examine how gate-resolved information affects charge inference, we consider the unbiased, biased, and antibiased decoder variants of Ref.~\cite{BarrattLearnability2022}. For a two-qubit gate \(U_i^{(t)}\), we define the local hopping probability
\begin{equation}
h\!\left(U_i^{(t)}\right)
=
\left|\langle 01|U_i^{(t)}|10\rangle\right|^2 .
\label{eq:hopping_amplitude}
\end{equation}
The corresponding two-site transfer matrix is
\begin{equation}
T_i^{(t)}
=
\begin{pmatrix}
1 & 0 & 0 & 0\\
0 & p_i^{(t)} & 1-p_i^{(t)} & 0\\
0 & 1-p_i^{(t)} & p_i^{(t)} & 0\\
0 & 0 & 0 & 1
\end{pmatrix},
\label{eq:sep_transfer_modes}
\end{equation}
where \(p_i^{(t)}\) is the decoder's effective probability that a single charge remains on the same side of the bond during the update. The three variants are
\begin{equation}
p_i^{(t)}
=
\begin{cases}
\tfrac{1}{2}, & \text{unbiased},\\
1-h\!\left(U_i^{(t)}\right), & \text{biased},\\
h\!\left(U_i^{(t)}\right), & \text{antibiased}.
\end{cases}
\label{eq:decoder_modes}
\end{equation}
The unbiased decoder ignores gate-specific hopping information and uses the Haar-averaged value. The biased decoder incorporates the hopping tendency associated with the actual gate realization and is therefore better matched to the physical circuit. The antibiased decoder intentionally uses the opposite local preference and serves as a controlled mismatched decoder. In all three cases, the monitored quantum circuit, weak-measurement channel, and physical record \(\mathcal M\) are identical; only the classical post-processing map from \(\mathcal M\) to \(\hat P_d(Q|\mathcal M)\) is changed.

We characterize charge learnability using prediction- and posterior-based diagnostics. The accuracy is the probability that the maximum-posterior prediction equals the true label. For a trajectory generated in sector \(Q^\ast\), the posterior probability assigned to the correct label is
\begin{equation}
P_{\mathrm{corr}}(\mathcal M)
=
\hat P_d(Q^\ast|\mathcal M).
\label{eq:pcorr}
\end{equation}
The posterior entropy is
\begin{equation}
S(\mathcal M)
=
-\sum_{Q=Q_0,Q_1}
\hat P_d(Q|\mathcal M)
\log_2 \hat P_d(Q|\mathcal M),
\label{eq:posterior_entropy}
\end{equation}
and the corresponding centered Binder ratio is
\begin{equation}
B_S
=
1-
\frac{
\left\langle
\left(S-\langle S\rangle\right)^4
\right\rangle
}{
3\left\langle
\left(S-\langle S\rangle\right)^2
\right\rangle^2
},
\label{eq:binder_ratio}
\end{equation}
where $\langle\cdot\rangle$ denotes an average over independent trajectories. 
This follows the posterior-entropy analysis of Ref.~\cite{BarrattLearnability2022}, where the entropy distribution provides a finite-size diagnostic of the learnability transition. The posterior entropy is label blind: it measures posterior sharpness but does not distinguish a sharp correct posterior from a sharp incorrect one. This distinction is important for deliberately mismatched decoders.

To make the diagnostic explicitly label-sensitive, we also compute the sample cross entropy
\begin{equation}
C(\mathcal M,Q^\ast)
=-\log_2 P_{\mathrm{corr}}(\mathcal M).
\label{eq:cross_entropy}
\end{equation}
The mean \(\langle C\rangle\) is the average log loss assigned to the true label, while \(\operatorname{Var}(C)\) measures trajectory-to-trajectory fluctuations in that loss and can serve as a finite-size transition diagnostic. Peaks in \(\operatorname{Var}(C)\) identify regimes of maximal sample-to-sample variation. Unlike the posterior-entropy Binder ratio, the cross entropy directly penalizes confident incorrect predictions, making the two diagnostics complementary when comparing decoder variants.

\subsection{Informational power of measurements}
Averaging over the measurement outcomes gives the local quantum channel
\begin{align}
\Phi_{q,\eta}(\rho)
&=(1-q)\rho+q\sum_{s=\pm}M_s(\eta)\rho M_s^\dagger(\eta)\nonumber\\
&=\left(1-\frac{q\eta^2}{1+\eta^2}\right)\rho
+\frac{q\eta^2}{1+\eta^2}Z\rho Z.
\label{eq:averaged_channel}
\end{align}
This channel is characterized by the ensemble dephasing parameter $q\eta^2/(1+\eta^2)$. Although useful for ensemble-averaged dynamics, this parameter does not quantify the information carried by an individual readout.

We therefore use the informational power of a measurement, introduced in Ref.~\cite{Arno_PRA_2011} and later applied to monitored quantum circuits~\cite{IppolitiKhemani2024Shadows}, to quantify the information available from a single readout.
The information extracted by a POVM $\Pi=\{\Pi_m\}$ from a state
ensemble $\mathcal{E}=\{p_i,\rho_i\}$ is quantified by the mutual
information between the input label $i$ and the measurement outcome
$m$. The corresponding joint and marginal probabilities are
\begin{equation}
p_{i,m}
=
p_i\,\mathrm{Tr}\!\left(\rho_i\Pi_m\right),
\qquad
p_m=\sum_i p_{i,m}.
\end{equation}
The mutual information, expressed in bits, is
\begin{equation}
I(\mathcal{E};\Pi)
=
\sum_{i,m}p_{i,m}
\log_2\frac{p_{i,m}}{p_i p_m}.
\label{eq:mutual_information}
\end{equation}
The single-readout informational power is the maximum mutual information over all input ensembles,
\begin{equation}
I_{\mathrm{read}}(\Pi)
\equiv
\max_{\mathcal{E}} I(\mathcal{E};\Pi).
\label{eq:informational_power}
\end{equation}
For the single-qubit weak measurement in Eq.~\eqref{eq:kraus_compact}, the POVM elements $\Pi_{\pm}=M_{\pm}^{\dagger}M_{\pm}$ are diagonal in the local $Z$ basis. The maximizing ensemble therefore consists of the two $Z$ eigenstates $\{\ket{z}:z=\pm1\}$ with equal prior probabilities. The conditional
probability of obtaining outcome $s=\pm1$ for input $\ket{z}$ is
\begin{equation}
p(s|z)
=
\bra{z}\Pi_s\ket{z}
=
\frac{(1+s z\eta)^2}{2(1+\eta^2)}.
\end{equation}
The readout therefore reduces to a binary symmetric channel: the
outcome correctly identifies $z$ with probability
$(1+\eta)^2/[2(1+\eta^2)]$ and gives the opposite result with error
probability
\begin{equation}
\epsilon(\eta)
=p(s=-z|z)
=\frac{(1-\eta)^2}{2(1+\eta^2)}.
\label{eq:weak_error_probability}
\end{equation}
Denoting the input and outcome random variables by $Z$ and $S$, equal priors give $H(S)=1$ and $H(S|Z)=h_2[\epsilon(\eta)]$, where $h_2(x)=-x\log_2x-(1-x)\log_2(1-x)$.
The single-readout informational power is therefore
\begin{equation}
I_{\mathrm{read}}(\eta)
=H(S)-H(S|Z)
=1-h_2[\epsilon(\eta)].
\label{eq:Iread_eta}
\end{equation}

Including the measurement probability \(q\) adds a no-measurement outcome that is independent of the local charge and therefore carries no information. We define the resulting local single-measurement informational power as
\begin{equation}
I_{\mathrm{loc}}(q,\eta)=q\,I_{\mathrm{read}}(\eta),
\label{eq:Iloc}
\end{equation}
which we refer to below as the local informational power. For projective measurements, \(I_{\mathrm{read}}(1)=1\), so \(I_{\mathrm{loc}}(q,1)=q\).

We test whether the finite-size learnability transition in the \((q,\eta)\) plane approximately follows constant \(I_{\mathrm{loc}}\). A projective protocol with measurement probability \(q_{\mathrm{s}}\) and a weak-measurement protocol \((q,\eta)\) are locally information matched when $q_{\mathrm{s}}=q\,I_{\mathrm{read}}(\eta)$.
This condition matches the local informational power, not the full monitored quantum channels. We test it using finite-size transition signatures along representative constant-\(I_{\mathrm{loc}}\) curves and across the three decoder variants defined above. 

\section{Phase diagram of charge learnability}
~\label{SecIII}
In this section, we first establish the learnability transition for deterministic weak measurements, then test the constant-\(I_{\mathrm{loc}}\) organization of the phase boundary and compare the three decoder variants.

Unless stated otherwise, the numerical results in this section use open chains of sizes \(L=6,8,10\) and a total circuit depth
\begin{equation}
T=2L,
\label{eq:numerical_depth}
\end{equation}
corresponding to 12, 16, and 20 alternating brickwork unitary layers, respectively. Each unitary layer is followed by a measurement layer, in which a local weak measurement of strength \(\eta\) is independently applied at every site with probability \(q\). For every parameter point \((L,q,\eta)\), we average over \(8\times10^4\) independent monitored trajectories, with equal numbers drawn from the two charge sectors \(Q_0=L/2\) and \(Q_1=L/2-1\). The random \(U(1)\)-symmetric gates and measurement outcomes are independently resampled for every trajectory. The physical evolution and SEP likelihood contractions are evaluated using MPS/TEBD methods with truncation cutoff \(10^{-8}\).

\revise{The dominant total numerical cost arises from the large trajectory ensembles together with the scans over measurement parameters and decoder variants. In particular, the Binder ratio involves fourth moments and \(\operatorname{Var}(C)\) is a fluctuation-based diagnostic, so stable estimates benefit from large trajectory ensembles. Accordingly, the present work focuses on robust finite-size transition signatures rather than thermodynamic critical scaling.}

\revise{As a representative sampling-convergence check, we subsampled the existing unbiased \(q=1\) weak-measurement data at \(N_{\rm traj}=10^4\), \(2\times10^4\), and \(4\times10^4\), and compared the resulting transition estimates with the full \(8\times10^4\)-trajectory ensemble. The Binder-crossing and \(\operatorname{Var}(C)\)-peak locations remain stable while their statistical uncertainties decrease, indicating that the reported finite-size estimates are not dominated by trajectory undersampling.} \revise{Statistical uncertainties are estimated from the finite trajectory ensemble. Error bars are shown for the cross-entropy variance \(\operatorname{Var}(C)\) and denote one standard error.}

\subsection{Learnability transition under probabilistic weak measurement}
We first examine whether the learnability transition persists when the projective measurements of the original problem are replaced by weak measurements with independently tunable measurement probability $q$ and measurement strength $\eta$. Our numerical results show that it does. 

\begin{figure*}[t]
\centering
\includegraphics[width=0.98\textwidth]{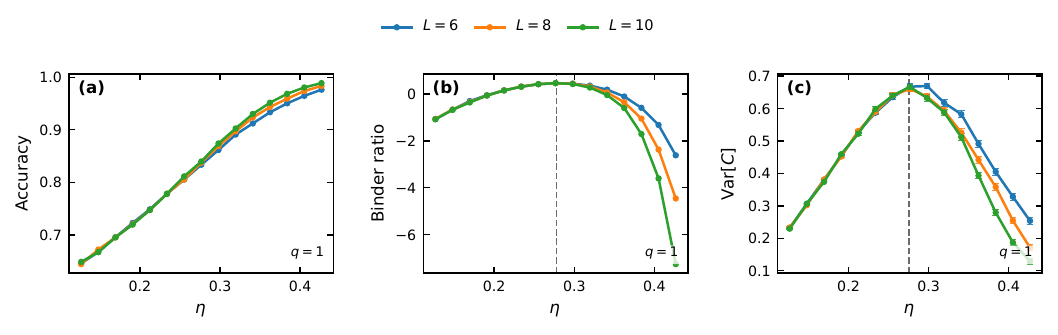}
\vspace{-0.7em}
\caption{
Finite-size diagnostics for the unbiased decoder at measurement probability $q=1$, with measurement strength $\eta$ varied along the horizontal axis. The representative condition $I_{\mathrm{loc}}\simeq0.20$ predicts $\eta_0\simeq0.277$. (a) Decoding accuracy. (b) Binder ratio of the binary posterior entropy. (c) Cross-entropy variance $\operatorname{Var}(C)$. The vertical dashed line marks the selected Binder crossing or variance peak. \revise{Error bars are shown only for \(\operatorname{Var}(C)\) in panel (c) and denote one standard error. Around the transition region, the one-standard-error uncertainties of the accuracy and Binder-ratio data are of order \(10^{-3}\) and are omitted for clarity.}
}
\label{fig:q1_eta_scan}
\end{figure*}

We begin with deterministic weak measurements, $q=1$, which isolate the dependence on measurement strength. This cut tests whether the transition occurs at the same local informational power as in the probabilistic projective case. 
As shown in Fig.~\ref{fig:q1_eta_scan}, for the $q=1$ weak-measurement cut, the decoding accuracy increases as the measurement strength $\eta$ increases, while the Binder-ratio and cross-entropy diagnostics develop transition-like features near the predicted constant-\(I_{\mathrm{loc}}\) point. 
For the representative line $I_{\mathrm{loc}}\simeq 0.2$, the matching condition $I_{\mathrm{loc}}=qI_{\mathrm{read}}(\eta)$ gives $\eta_0\simeq0.277$ at $q=1$. 
The posterior-entropy Binder ratio exhibits finite-size crossings near this predicted value, and $\operatorname{Var}(C)$ peaks in the same parameter regime.

%The coexistence of these signatures demonstrates that weak measurements do not destroy the learnability physics identified in the projective problem~\cite{BarrattLearnability2022}. 
%The physical interpretation is straightforward can be follows. 
%Increasing $\eta$ at fixed $q$ increases the amount of label-relevant information injected into the measurement record, and the decoder response changes accordingly. The persistence of both Binder-ratio crossings and cross-entropy fluctuations under this generalization shows that the learnability transition exists not only for ideal projective readout.

\begin{figure}[t]
\centering
\includegraphics[width=0.98\columnwidth]{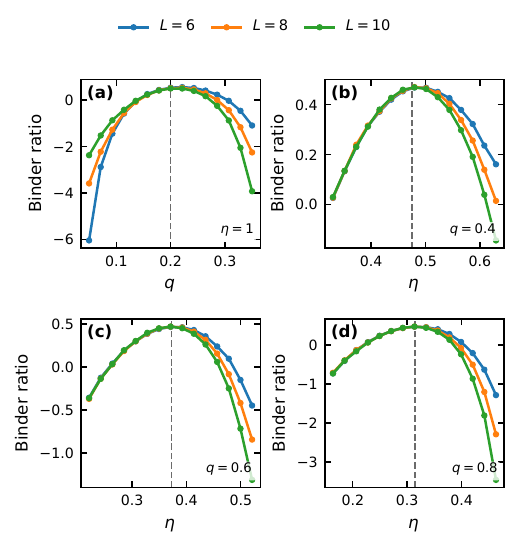}
\vspace{-0.7em}
\caption{
Binder-ratio tests for the unbiased decoder along the representative line $I_{\mathrm{loc}}\simeq0.20$. (a) Measurement-probability scan at the projective limit $\eta=1$, with predicted transition $q_0=0.20$. (b)--(d) Measurement-strength scans at fixed $q=0.40$, $0.60$, and $0.80$, with predicted transitions $\eta_0=0.480$, $0.371$, and $0.314$, respectively. The vertical dashed lines mark the selected finite-size Binder crossings. \revise{Around the selected crossings, the one-standard-error uncertainties of the Binder ratio are of order \(10^{-3}\) and are omitted for clarity.}
}
\label{fig:cross_tests_unbiased_endpoints}
\end{figure}

The agreement between the probabilistic projective and deterministic weak-measurement cuts motivates extending the one-parameter learnability transition to a boundary in the two-dimensional parameter space $(q,\eta)$. We test whether its finite-size locations approximately follow constant $I_{\mathrm{loc}}$. 
The candidate organizing variable is the local informational power $I_{\mathrm{loc}}(q,\eta)=qI_{\mathrm{read}}(\eta)$, which combines the measurement probability with the single-readout informational power. 
As shown in Fig.~\ref{fig:cross_tests_unbiased_endpoints}, we test this prediction directly by following the representative constant-\(I_{\mathrm{loc}}\) line $I_{\mathrm{loc}}\simeq0.2$. 
The projective limit corresponds to a fixed-$\eta$ scan at $\eta=1$, with the predicted transition point $q_0=0.20$. 
The remaining panels show fixed-$q$ scans in $\eta$, with $q=0.40$, $0.60$, and $0.80$, for which the predicted transition points are $\eta_0=0.480$, $0.371$, and $0.314$, respectively.

The observed Binder-ratio crossings occur close to these predicted values. 
Together with the $q=1$ cut in Fig.~\ref{fig:q1_eta_scan}, this provides a direct finite-size test of the constant-$I_{\mathrm{loc}}$ organization. 
We also find that the corresponding $\operatorname{Var}(C)$ diagnostics give compatible transition estimates, although the main evidence displayed here is the Binder-ratio crossing structure.

This result has a direct interpretation. Learnability depends jointly on the measurement probability and strength through the local informational power supplied to the record. Protocols with different $(q,\eta)$ can therefore exhibit similar decoding behavior when they have comparable $I_{\mathrm{loc}}$. The agreement between the observed crossings and predicted locations supports $I_{\mathrm{loc}}$ as an operational link between projective and weak measurements.

\begin{figure}[t]
\centering
\includegraphics[width=0.98\columnwidth]{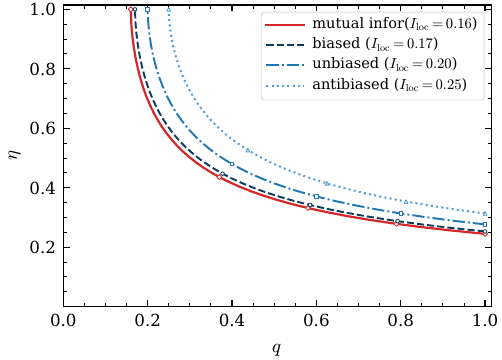}
\vspace{-0.5em}
\caption{
Decoder-independent and decoder-dependent charge-learnability boundaries in the plane of measurement probability $q$ and measurement strength $\eta$. The red solid curve denotes the decoder-independent finite-size boundary inferred from the exact record--label mutual information, $I_{\mathrm{loc}}\simeq0.16$. The three blue curves denote the biased, unbiased, and antibiased transition estimates, with $I_{\mathrm{loc}}\simeq0.17$, $0.20$, and $0.25$, respectively. Open markers indicate representative finite-size scan points. All curves satisfy $qI_{\mathrm{read}}(\eta)=I_{\mathrm{loc}}$.
}
\label{fig:Iloc_phase_overview}
\end{figure}

The resulting transition boundaries are summarized in Fig.~\ref{fig:Iloc_phase_overview}. In addition to the three decoder-dependent boundaries, we include the decoder-independent finite-size boundary inferred from the exact record--label mutual information. The red solid curve corresponds to \(I_{\mathrm{loc}}\simeq0.16\), while the biased, unbiased, and antibiased decoder boundaries correspond to \(I_{\mathrm{loc}}\simeq0.17\), \(0.20\), and \(0.25\), respectively. The markers denote representative scan points used in the finite-size analyses. The scans in Figs.~\ref{fig:q1_eta_scan} and~\ref{fig:cross_tests_unbiased_endpoints} are cuts through these constant-\(I_{\mathrm{loc}}\) curves.

\subsection{Decoder dependence of the transition estimates}
\begin{figure*}[t]
\centering
\includegraphics[width=0.98\textwidth]{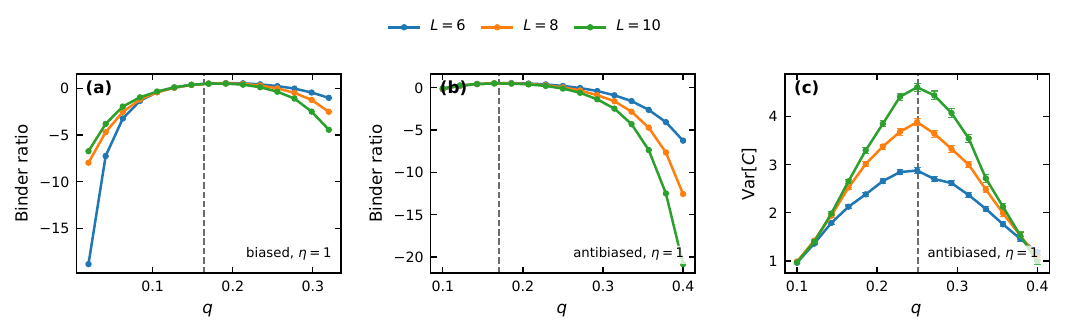}
\vspace{-0.7em}
\caption{
Finite-size decoder diagnostics in the projective limit $\eta=1$, with the measurement probability $q$ varied along the horizontal axis. (a) For the biased decoder, the posterior-entropy Binder ratio locates the transition near $q_0\simeq0.17$. (b) For the antibiased decoder, the same label-blind diagnostic gives an apparent estimate near $q_0\simeq0.17$. (c) The label-sensitive cross-entropy variance $\operatorname{Var}(C)$ instead peaks near $q_0\simeq0.25$. Vertical dashed lines mark the selected crossings or peak. \revise{Error bars are shown only for \(\operatorname{Var}(C)\) in panel (c) and denote one standard error. Around the Binder crossings, the one-standard-error Binder-ratio uncertainties are of order \(10^{-3}\) and are omitted for clarity.}
}
\label{fig:strong_decoders_binder}
\end{figure*}

\begin{figure}[t]
\centering
\includegraphics[width=0.98\columnwidth]{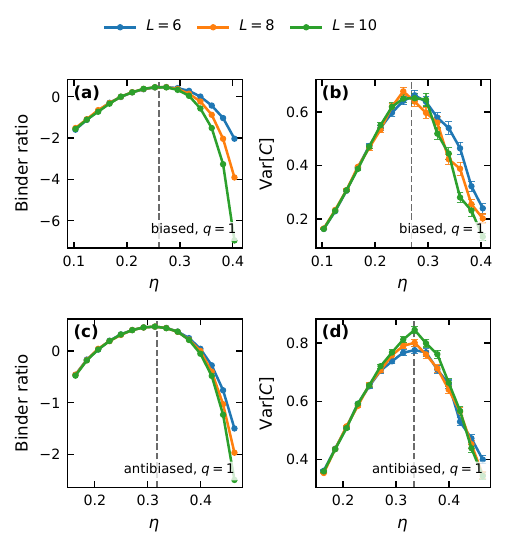}
\vspace{-0.7em}
\caption{
Finite-size decoder diagnostics for deterministic weak measurements at $q=1$, with the measurement strength $\eta$ varied along the horizontal axis. The top row shows the biased decoder, with transition features near $\eta_0\simeq0.253$ ($I_{\mathrm{loc}}\simeq0.17$); the bottom row shows the antibiased decoder, with features near $\eta_0\simeq0.314$ ($I_{\mathrm{loc}}\simeq0.25$). The left column displays the posterior-entropy Binder ratio and the right column the cross-entropy variance $\operatorname{Var}(C)$. Vertical dashed lines mark the selected crossings or peaks. \revise{Error bars in the right column are shown for \(\operatorname{Var}(C)\) and denote one standard error. Around the Binder crossings, the one-standard-error Binder-ratio uncertainties are of order \(10^{-3}\) and are omitted for clarity.}
}
\label{fig:decoder_binder_ce_comparison}
\end{figure}

We next compare the biased, unbiased, and antibiased decoder variants. Their expected performance is ordered from the deliberately mismatched antibiased decoder, through the unbiased decoder, to the gate-matched biased decoder. Posterior entropy alone cannot distinguish a sharp correct inference from a sharp incorrect one, so we supplement its Binder ratio with the label-sensitive cross entropy. Figures~\ref{fig:strong_decoders_binder} and~\ref{fig:decoder_binder_ce_comparison} examine two complementary cuts of the phase diagram: a measurement-probability scan in the projective limit $\eta=1$, and a measurement-strength scan at $q=1$.

We first fix $\eta=1$ and vary the measurement probability $q$, as shown in Fig.~\ref{fig:strong_decoders_binder}. For the biased decoder, the posterior-entropy Binder ratio locates the transition near $q_0\simeq0.17$. For the antibiased decoder, the same label-blind diagnostic gives an apparent estimate near $q_0\simeq0.17$, whereas the cross-entropy variance peaks near $q_0\simeq0.25$. This discrepancy demonstrates why a label-sensitive diagnostic is required for a deliberately mismatched decoder.

We then fix the measurement probability at $q=1$ and vary the measurement strength $\eta$, as shown in Fig.~\ref{fig:decoder_binder_ce_comparison}. For the biased decoder, both the posterior-entropy Binder ratio and $\operatorname{Var}(C)$ identify a transition near $\eta_0\simeq0.253$, corresponding to $I_{\mathrm{loc}}\simeq0.17$. For the antibiased decoder, the corresponding features occur near $\eta_0\simeq0.314$, corresponding to $I_{\mathrm{loc}}\simeq0.25$. Thus the deterministic weak-measurement scan is consistent with the cross-entropy scale selected in the projective limit.

Taken together, Figs.~\ref{fig:strong_decoders_binder} and~\ref{fig:decoder_binder_ce_comparison} show that the low-$q$ Binder crossing of the antibiased decoder is not stable under a label-sensitive diagnostic. Both cuts give a consistent biased-decoder scale, $I_{\mathrm{loc}}\simeq0.17$. For the antibiased decoder, the cross-entropy variance selects $I_{\mathrm{loc}}\simeq0.25$ in the projective limit, and the $q=1$ measurement-strength scan supports the same scale.

The distinction follows from what the two diagnostics measure. The Binder ratio probes posterior sharpness and therefore does not distinguish ``sharp and correct'' from ``sharp but wrong.'' This is especially relevant for a deliberately mismatched decoder, which can become confidently incorrect. By contrast, the cross entropy depends directly on the probability assigned to the true label. Its variance is therefore a more appropriate transition diagnostic when decoder mismatch complicates the posterior-entropy distribution. Consequently, transition estimates from mismatched decoders should be checked against a label-sensitive diagnostic.

\section{Mutual information as a limit on charge learnability} \label{sec:mi}
The projective--weak correspondence and decoder comparison motivate a decoder-independent characterization based on the mutual information between the hidden label and the measurement record. Let \(Y\in\{0,1\}\) denote the balanced binary label associated with sectors \(Q_0\) and \(Q_1\), and let \(\mathcal M\) denote the complete spacetime measurement record. The record--label mutual information is~\cite{CoverThomas2006}
\begin{align}
I_{\mathrm{true}}
&\equiv I(Y;\mathcal M) \nonumber\\
&=
\sum_{y,\mathcal M}
P(y,\mathcal M)
\log_2
\frac{P(y,\mathcal M)}{P(y)P(\mathcal M)}
\nonumber\\
&=H(Y)-H(Y|\mathcal M).
\label{eq:Itrue}
\end{align}
For the balanced binary task, \(H(Y)=1\) bit.

For each realized record, we evaluate its exact likelihood under both candidate sectors. The initial ensemble in sector \(Q\) is uniform over all computational-basis states with total charge \(Q\), and the probability of observing record $\mathcal M$ is
\begin{equation}
P(\mathcal M|Q)
=
\frac{1}{d_Q}
\sum_{\boldsymbol z:\,|\boldsymbol z|=Q}
\left\|
\mathcal K_{\mathcal M}|\boldsymbol z\rangle
\right\|^2,
\qquad
d_Q=\binom{L}{Q},
\label{eq:exact_record_likelihood}
\end{equation}
where \(\mathcal K_{\mathcal M}\) is the time-ordered product of the charge-conserving unitary gates and the measurement Kraus operators specified by the record. The factors associated with the occurrence or absence of measurements are common to the two candidate sectors and therefore cancel in the posterior likelihood ratio. Numerically, Eq.~\eqref{eq:exact_record_likelihood} is evaluated by propagating every computational-basis state in the candidate sector and averaging the corresponding probabilities of the observed record.

For equal prior probabilities, Bayes' rule gives the exact posterior
\begin{equation}
P(Q_a|\mathcal M)
=
\frac{P(\mathcal M|Q_a)}
{P(\mathcal M|Q_0)+P(\mathcal M|Q_1)},
\qquad a=0,1.
\label{eq:true_posterior}
\end{equation}
Writing \(p_{\mathcal M}=P(Q_1|\mathcal M)\), the information associated with an individual trajectory is
\begin{equation}
i_{\mathrm{true}}(\mathcal M)
=
1-h_2(p_{\mathcal M}),
\label{eq:itraj}
\end{equation}
and averaging over trajectories gives
\begin{equation}
I_{\mathrm{true}}
=
\left\langle i_{\mathrm{true}}(\mathcal M)\right\rangle_{\mathcal M}
=
1-\left\langle h_2(p_{\mathcal M})\right\rangle_{\mathcal M}.
\label{eq:Itrue_from_binary}
\end{equation}
This exact posterior is determined directly from the physical record probabilities and is therefore independent of the chosen SEP decoder variant.

To obtain a finite-size transition estimate, we apply the centered Binder-ratio analysis of Eq.~\eqref{eq:binder_ratio}, replacing the decoder posterior entropy \(S(\mathcal M)\) by the trajectory-resolved information \(i_{\mathrm{true}}(\mathcal M)\). We denote the resulting Binder ratio by \(B_I\). The sample-to-sample variance \(\operatorname{Var}(i_{\mathrm{true}})\) is used as a complementary diagnostic: its maximum gives a size-dependent pseudocritical point, while crossings of \(B_I\) between different system sizes estimate the transition boundary. Because \(i_{\mathrm{true}}=1-S_{\mathrm{true}}\), the centered Binder ratio of \(i_{\mathrm{true}}\) is identical to that of the exact posterior entropy.

The centered Binder crossings and variance peaks consistently place the decoder-independent finite-size transition near \(I_{\mathrm{loc}}\simeq0.16\). We therefore use the corresponding constant-\(I_{\mathrm{loc}}\) curve as the mutual-information boundary shown by the red solid line in Fig.~\ref{fig:Iloc_phase_overview}. The open markers on this and the decoder-dependent curves indicate the representative parameter points at which the finite-size scans were performed.

\begin{figure}[t]
\centering
\includegraphics[width=0.95\columnwidth]{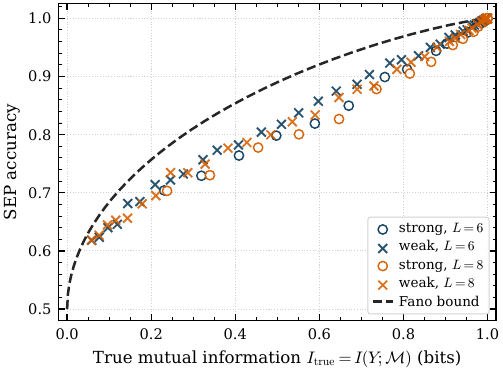}
\caption{
Projective--weak comparison using the exact record--label mutual information. The unbiased SEP decoding accuracy is plotted against the exact record--label mutual information $I_{\mathrm{true}}$ for projective measurements and deterministic weak measurements. Exact-posterior evaluation is performed for $L=6,8$ at depth $T=2L$, using $10^4$ independent trajectories per parameter point with balanced charge labels. \revise{The dashed curve shows the decoder-independent binary Fano upper bound on the achievable classification accuracy~\cite{CoverThomas2006,Fano1961}; it represents an information-theoretic envelope rather than a prediction for the SEP decoder.}
}
\label{fig:mi_collapse}
\end{figure}

The exact-posterior calculations use the same alternating brickwork architecture and depth $T=2L$, but are restricted to smaller sizes because Eq.~\eqref{eq:exact_record_likelihood} requires propagation over every basis state in each candidate charge sector. For the projective--weak comparison in Fig.~\ref{fig:mi_collapse}, we use $L=6,8$ and average over $10^4$ independent trajectories for each parameter point, with equal sampling of the two charge labels.

Beyond locating the transition boundary, we examine how decoder performance depends on the total information carried by the record. For the balanced binary task, $H(Y)=1$ and therefore $H(Y|\mathcal M)=1-I_{\mathrm{true}}$. The binary Fano inequality gives
\begin{equation}
1-I_{\mathrm{true}}\le h_2(P_{\mathrm e}),
\end{equation}
where $P_{\mathrm e}$ is the minimum classification-error probability. Taking the inverse of $h_2$ on the interval $[0,1/2]$, the achievable accuracy satisfies
\begin{equation}
P_{\mathrm{acc}}\le 1-h_2^{-1}(1-I_{\mathrm{true}}).
\label{eq:fano_accuracy_bound}
\end{equation}
As shown in Fig.~\ref{fig:mi_collapse}, the projective- and weak-measurement data for the unbiased decoder approximately collapse onto a common increasing curve when plotted against $I_{\mathrm{true}}$. Data from different scan paths and system sizes overlap within the finite-size scatter, supporting the interpretation that $q$ and $\eta$ primarily tune the information carried by the measurement record. The SEP accuracy remains below the binary Fano upper bound, as required for any decoder using the same record.

\revise{We emphasize that the Fano curve is a decoder-independent information-theoretic upper envelope rather than a prediction for the SEP decoder itself. Its curvature follows from the nonlinear inverse binary-entropy function in Eq.~\eqref{eq:fano_accuracy_bound}. The SEP decoder is a specific approximate inference scheme and is not expected to saturate this bound. Consequently, its accuracy can remain below the Fano envelope and appear approximately linear over the finite range of \(I_{\mathrm{true}}\) explored here without any inconsistency with the Fano inequality.}

This record--label mutual-information perspective further supports a decoder-independent interpretation of the transition. Different decoders need not have identical finite-size performance: the biased decoder extracts a larger fraction of the available information than the unbiased decoder, whereas the antibiased decoder extracts a smaller fraction because of its deliberately mismatched update rule. Thus decoder dependence concerns extraction efficiency rather than the information contained in the physical record. The quantity \(I_{\mathrm{loc}}(q,\eta)\) characterizes the local informational power supplied at each monitored spacetime point, while \(I_{\mathrm{true}}\) quantifies the total information carried by the complete record about the hidden label. Together, they provide an information-theoretic description of the projective--weak correspondence.

\section{Conclusions}
~\label{SecIV}
We have extended the learnability problem of a conserved global charge from probabilistic projective measurements to probabilistic weak measurements. Our results show that the learnability transition persists beyond ideal projective readout: increasing either the measurement probability or the measurement strength drives the record from an information-poor regime to one from which the hidden charge can be reliably inferred.

We find that the finite-size transition boundary is organized primarily by the local informational power of the measurement. This provides an approximate projective--weak correspondence: probabilistic weak and projective monitoring protocols exhibit similar learnability when they supply comparable local information to the measurement record. We have also introduced cross entropy as a label-sensitive diagnostic that clearly distinguishes unbiased, biased, and antibiased decoders, avoiding misleading transition estimates produced by mismatched decoding rules. The record--label mutual information analysis further supports a decoder-independent interpretation of the transition. Its crossings and fluctuation peaks identify an intrinsic information scale below the apparent transition scales of approximate decoders. Projective- and weak-measurement data follow a common relation between decoding accuracy and the mutual information carried by the full record. This indicates that mutual information sets the information fundamentally available for charge inference, while the decoder determines how efficiently it is recovered.

Our work identifies the informational power of each local measurement as a key ingredient, although spacetime correlations within the measurement record remain essential to the learnability transition. The proposed criterion remains conjectural and is supported mainly by finite-size numerics, but it motivates a broader information-based description of monitored dynamics. Such a framework should clarify the interplay among local measurement informational power, the scrambling and redistribution of information under different circuit architectures, decoder performance, and the fundamental mutual information between the hidden label and the complete record, which remains for further investigation.

\begin{acknowledgments}
This work was supported by the National Natural Science Foundation of China (Grant No.12375013, No.12547109), and Guangdong Provincial Quantum Science Strategic Initiative (Grant No. GDZX2503008).   
\end{acknowledgments}

% Optional appendix content can be inserted here.
% For example: additional scans, finite-size data tables,
% supplementary decoder-comparison figures, and calibration checks.

%\bibliographystyle{apsrev4-2}
\bibliography{references_44_refs}

\end{document}